\documentclass[apj]{emulateapj}
\usepackage{times}
\usepackage{graphicx}
\usepackage{amssymb}
\usepackage{multirow}
\usepackage{color}
\usepackage{wrapfig}

\usepackage{float}

\usepackage{epstopdf}

\newcommand{\fermi }{{\it Fermi}}

\newif\ifAMStwofonts
\AMStwofontstrue

\makeatletter

\newcommand{\Rmnum}[1]{\expandafter\@slowromancap\romannumeral #1@}
\makeatother

\shorttitle{Annular Gap Model for the Vela Pulsar}

\shortauthors{Du et al. 2011}

\begin{document}





\title{Gamma-ray emission from the Vela pulsar modeled with the
  Annular Gap and Core Gap}

\author{Y.~ J.~ Du\altaffilmark{1},~
J.~ L.~ Han\altaffilmark{1},~
G.~ J.~ Qiao\altaffilmark{2}
~and~
C.~ K.~ Chou\altaffilmark{1}}

\altaffiltext{1}{National Astronomical Observatories, Chinese Academy
  of Sciences, Jia 20 Datun Road, Beijing 100012, China}
\altaffiltext{2}{School of Physics, Peking University, Beijing 100871,
  China}



\begin{abstract}

The Vela pulsar represents a distinct group of $\gamma$-ray pulsars.
{\fermi} $\gamma$-ray observations reveal that it has two sharp peaks
(P1 and P2) in the light curve with a phase separation of 0.42 and a
third peak (P3) in the bridge. The location and intensity of P3 are
energy-dependent. We use the 3D magnetospheric model for the annular
gap and core gap to simulate the $\gamma$-ray light curves,
phase-averaged and phase-resolved spectra.
We found that the acceleration electric field along a field line in
the annular gap region decreases with heights. The emission at high
energy GeV band is originated from the curvature radiation of
accelerated primary particles, while the synchrotron radiation from
secondary particles have some contributions to low energy $\gamma$-ray
band ($0.1 - 0.3$~GeV).
The $\gamma$-ray light curve peaks P1 and P2 are generated in the
annular gap region near the altitude of null charge surface, whereas
P3 and the bridge emission is generated in the core gap region. The
intensity and location of P3 at different energy bands depend on the
emission altitudes.
The radio emission from the Vela pulsar should be generated in a
high-altitude narrow regions of the annular gap, which leads to a
radio phase lag of $\sim$ 0.13 prior to the first $\gamma$-ray
peak.

\end{abstract}

\keywords{Pulsars: general -- Gamma rays: stars -- radiation
mechanisms: non-thermal -- Pulsars: individual (PSR J0835-4510)}

\section{Introduction}

The Vela pulsar is the brightest point source in the $\gamma$-ray
sky. The Vela pulsar at a distance of $d=287_{-17}^{+19}$\,pc
\citep{2003ApJ...596.1137D} has a spin period of $P=89.3$\,ms,
characteristic age $\tau_{\rm c}=11$\,kyr, magnetic field $B =
3.38\times 10^{12}$\,G, and the rotational energy loss rate
$\dot{E}=6.9\times 10^{36} \rm \,erg\,s^{-1}$
\citep{2005AJ....129.1993M}. It radiates multi-waveband pulsed
emission from radio to $\gamma$-ray which enables us to get
considerable insights of the magnetosphere activities. High energy
$\gamma$-ray emission takes away a significant fraction of the
spin-down luminosity \citep{1999ApJ...516..297T, 2001AIPC..558..103T}.
The pulsed $\gamma$-ray emission from the Vela pulsar was detected by
many instruments, e.g. SAS 2 \citep{1975ApJ...200L..79T}, COS B
\citep{1988A&A...204..117G}, the Energetic Gamma Ray Experiment
Telescope (EGRET, Kanbach et al. 1994; Fierro et al. 1998),
Astro-rivelatore Gamma a Immagini LEggero (AGILE, Pellizzoni et
al. 2009) and {\fermi} \citep{2009ApJ...696.1084A,
  2010ApJ...713..154A}. The $\gamma$-ray profile has two main sharp
peaks (P1 and P2) and a third peak (P3) in the bridge. The location
and intensity of P3 as well as the peak ratio (P1/P2) vary with energy
\citep{2010ApJ...713..154A, 2010ApJS..187..460A}.  Because of the
large $\dot{E}$, the Vela pulsar has a strong wind nebulae, from which
the unpulsed $\gamma$-ray photons was detected by AGILE
\citep{2010Sci...327..663P} and {\fermi} \citep{2010ApJ...713..146A}.

Theories for non-thermal high energy emission of pulsars are
significantly constrained by sensitive $\gamma$-ray observations by
the {\fermi} telescope.
Four physical or geometrical magnetospheric models have previously
been proposed to explain pulsed $\gamma$-ray emission of pulsars: the
polar cap model \citep{1994ApJ...429..325D, 1996ApJ...458..278D} in
which the emission region is generated near the neutron star surface,
the outer gap model \citep{1986ApJ...300..500C, 1986ApJ...300..522C,
  1995ApJ...438..314R, 1997ApJ...487..370Z, 2000ApJ...537..964C,
  2004ApJ...604..317Z, 2007ApJ...666.1165Z, 2008ApJ...688L..25H,
  2008ApJ...676..562T, 2009ApJ...699.1711L} in which the emission
region is generated near the light cylinder, the two-pole caustic
model or the slot gap model \citep{2003ApJ...598.1201D,
  2003ApJ...588..430M, 2004ApJ...606.1143M, 2008ApJ...680.1378H} in
which the emission region is generated along the last open field
lines, and the annular gap model \citep{2004ApJ...606L..49Q,
  2004ApJ...616L.127Q, 2007ChJAA...7..496Q, DQHLX+10} in which the
emission is generated near the null charge surface. The distinguishing
features of these models are different acceleration region for primary
particles and possible mechanisms to radiate high energy photons.
\citet{1995ApJ...438..314R} modeled the $\gamma$-ray and radio light
curves for the Vela pulsar with a larger viewing angle ($\zeta \sim
79^{\circ}$). In their outer gap model, the two $\gamma$-ray peaks are
generated from the outer gap of one pole, whereas the radio emission
is radiated from the other pole. However, the correlation of high
energy X-ray emission and the radio pulse shown by
\citet{2007ApJ...657..436L} is not consistent with this picture.
\citet{2003ApJ...598.1201D} used the two-pole caustic model to
simulate the $\gamma$-ray light curve for the Vela pulsar, which was
further revised by \citet{2009RAA.....9.1324Y} and
\cite{2010ApJ...709..605F} to explain the details of {\fermi} GeV
light curves. A bump appears in the bridge in the model for a large
inclination angle, but the width and the location of P3 were not well
modeled yet.

In this paper, we focus on the $\gamma$-ray light curves at differnt
bands and spectra of the Vela pulsar. In \S\,2, we introduce the
annular gap and core gap, and calculate the acceleration electric
field in the annular gap. In \S\,3, we model the multi-band light
curves using the annular gap model together with a core gap. The radio
emission region is identified and the radio lag prior to the first
$\gamma$-ray peak is explained. To model the Vela pulsar spectrum, we
also calculate the $\gamma$-ray phase-averaged and phase-resolved
spectra of both synchro-curvature radiation from the primary particles
and sychrotron radiation from the secondary particles. Conclusions and
discussions are presented in \S\,4.

\begin{figure}[tb]
\centering
\includegraphics[angle=0,scale=.47]{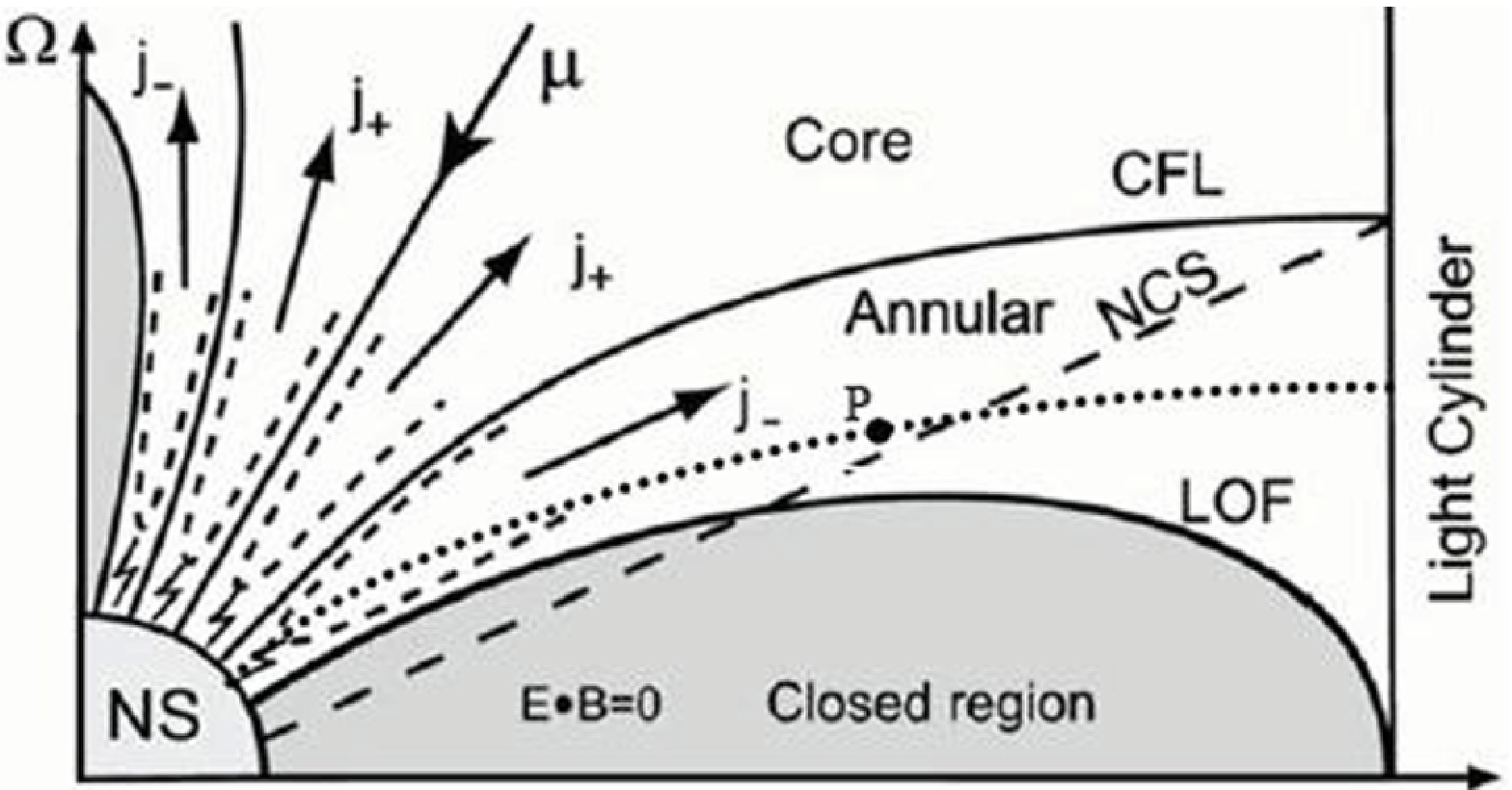}
\caption{Sketch for the annular gap and the core gap for particle
  acceleration. CFL stands for the critical field line across the
  intersection of the null charge surface and the light cylinder, NCS
  for the null charge surface, and LOF for the last open field line. P
  is the peak emission spot at a field line in the annular gap region,
  which is located between CFL and LOF. This figure is taken and
  revised from Figure 1 of
  \citet{2007ChJAA...7..496Q}. } \label{AG-sketch}
\end{figure}

\section{The annular gap and core gap}

\subsection{Formation of the Annular Gap and the Core Gap}

The open field line region of pulsar magnetosphere can be divided into
two parts by the critical field lines (see Figure~\ref{AG-sketch}).
The core region near the magnetic axis is defined by the critical
field lines. The annular region is located between the critical field
lines and the last open field lines. For an anti-parallel rotator the
radius of the core gap ($r_{\rm core}$) and the full polar cap region
($r_{\rm p}$) are $r_{\rm core} = (2/3)^{3/4}R(\Omega R/c)^{1/2}$ and
$r_{\rm p} = R(\Omega R/c)^{1/2}$, respectively \citep{RS75}, where
$R$ is the neutron star radius, $\Omega$ is the angular velocity
($\Omega=2\pi/P$, $P$ is the pulsar spin period). The radius of the
annular polar region therefore is $r_{\rm ann}=r_{\rm p}-r_{\rm core}
= 0.26 R(\Omega R/c)^{1/2}$. It is larger for pulsars with smaller
spin periods.

The annular acceleration region is negligible for older long period
pulsars, but very important for pulsars with a small period, e.g.,
millisecond pulsars and young pulsars. It extends from the pulsar
surface to the null charge surface or even beyond it (see
Figure~\ref{AG-sketch}). The annular gap has a sufficient thickness of
trans-field lines and a wide altitude range for particle
acceleration. In the annular gap model, the high energy emission is
generated in the vicinity of the null charge surface
\citep{DQHLX+10}. This leads to a fan-beam $\gamma$-ray emission
\citep{2007ChJAA...7..496Q}. The radiation from both the core gap and
the annular gap can be observed by one observer
\citep{2004ApJ...616L.127Q} if the inclination angle and the viewing
angle are suitable.

\subsection{Acceleration Electric Field}

The charged particles can not co-rotate with a neutron star near the
light cylinder and must escape from the magnetosphere. If particles
escape near the light cylinder, these particles have to be generated
and move out from the inner region to the outer region. This dynamic
process is always taking place, and a huge acceleration electric field
exists in the magnetosphere. To keep the whole system charge-free, the
neutron star surface must supply the charged particles to the
magnetosphere.

The annular gap and the core gap have particles with opposite sign
flowing, which can lead to the circuit closure in the whole
magnetosphere. The potential along the closed field lines and the
critical field lines are different \citep{XCQ06}. The parallel
electric fields ($E_{\parallel}$) in the annular gap and core gap
regions are opposite, as has been discussed by \citet{S71}. As a
result, $E_{\parallel}$ vanishes at the boundary (i.e. the critical
field lines) between the annular and the core regions and also along
the closed field lines. The positive and the negative charges are
accelerated from the core and the annular regions, respectively.

\begin{figure}
\centering
\includegraphics[angle=-90,scale=.58]{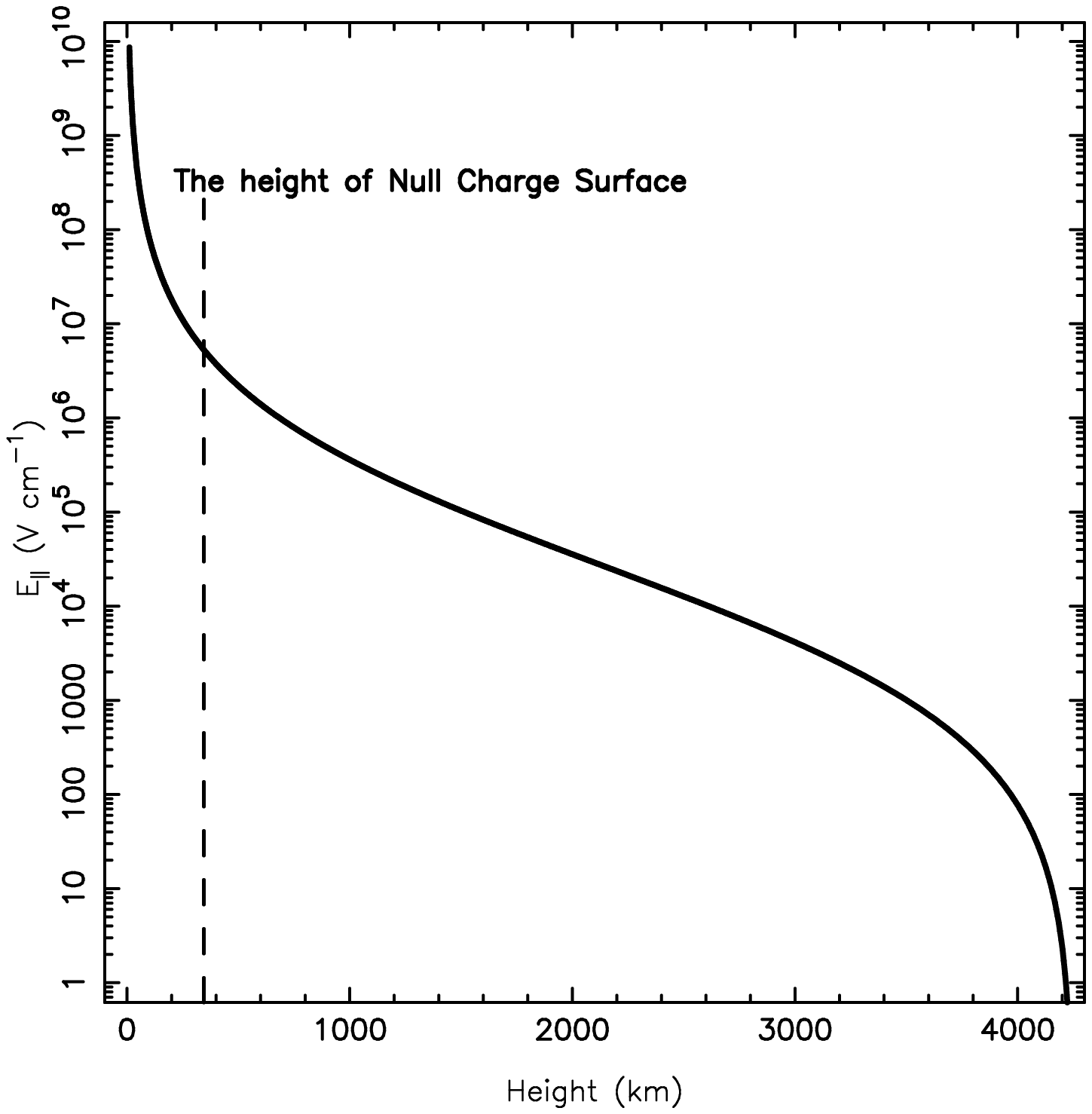}
\caption{For the Vela pulsar, the acceleration electric field is
  calculated along the field line marked as the dotted line with `P'
  in Figure~\ref{AG-sketch} with a magnetic azimuthal $\psi=0^\circ$
  and a magnetic inclination angle $\alpha=70^{\circ}$ in the annular
  gap model.} \label{ACC}
\end{figure}

We now consider a tiny magnetic tube in the annular gap region.  We
assume that the particles flow out at a radial distance about $r_{\rm
  out}\sim R_{\rm LC}=4.3\;10^3$~km, and that the charge density of
flowing-out particles $\rho_{\rm b}(r_{\rm out})$ is equal to the
local Goldreich-Julian (GJ) charge density $\rho_{\rm gj}(r_{\rm
  out})$ \citep{GJ69} at a radial distance of $r_{\rm out}$. For any
heights $r<r_{\rm out}$,
$\rho_{\rm b}(r) < \rho_{\rm gj}(r)$.
The acceleration electric field therefore exists along the field line,
and cannot vanish until approaching the height of $r_{\rm out}$.

For a static dipole magnetic field, the field components can be
described as
$\mathbf{B}_r=\frac{2\mu \cos\theta}{r^3}\mathbf{n}_r$
and
$\mathbf{B}_{\theta}=\frac{\mu \sin\theta}{r^3}\mathbf{n}_{\theta}$,
here $\theta$ is the zenith angle in magnetic field coordinate, and
$B_0$ is the surface magnetic field. Thus the magnetic field at a
height $r$ is
$B(r)=\frac{B_0 R^3}{2}\frac{\sqrt{3\cos^2\theta+1}}{r^3}$.
In the co-rotating frame, Poisson's equation is
\begin{equation}
\nabla \cdot \mathbf{E}=4\pi(\rho_{\rm b} - \rho_{\rm gj}). \label{pos}
\end{equation}
Because of the conservation laws of the particle number and magnetic
flux in the magnetic flux tube, the difference between the flowing
charge density and local GJ charge density at the radius $r$ can be
written as
\begin{equation}
\rho_{\rm b}(r)-\rho_{\rm gj}(r) =
-\frac{\Omega B(r)}{2 \pi c}(\cos\zeta_{\rm out}-\cos\zeta),
\label{rho}
\end{equation}
where $\Omega=2\pi/P$ is the angular velocity, $P$ is the rotation
period, and $\zeta$ (and $\zeta_{\rm out}$) are the angle between the
rotational axis and the $B$ field direction at $r$ (and $r_{\rm
  out}$). Wang et al. (2006) found
\begin{equation}
\cos\zeta
=\cos\alpha\cos\theta_{\mu}-\sin\alpha\sin\theta_{\mu}\cos\psi,
\label{zet}
\end{equation}
where $\psi$ and $\theta_{\mu}$ are the azimuthal angle and the
tangent angle (half beam angle) in the magnetic field coordinate,
respectively. Combining equations (\ref{pos}), (\ref{rho}) and
(\ref{zet}), we obtain
\begin{equation}
\nabla \cdot {\rm \bf E}
= -\frac{\Omega B_0 R^3}{c r^3}{\sqrt{3\cos^2\theta+1}}
(\cos\zeta_{\rm out}-\cos\zeta).
\label{eee}
\end{equation}
Substituting $\cot\theta_{\mu}=\frac{2\cot^2\theta-1}{3\cot\theta}$
\citep{1998A&A...333..172Q} and $ds=\sqrt{(rd\theta)^2+(dr)^2}$ into
equations (\ref{zet}) and (\ref{eee}), We can solve the equation for
$\nabla \cdot \mathbf{E}$, and calculate the electric field
$E_{\parallel}$ along a magnetic filed line for $\psi=0^\circ$, as
shown in Figure \ref{ACC} for the Vela pulsar. The electric field is
huge in the inner region of annular gap and drops quickly when $r
 \sim R_{\rm LC}$.

\section{Modeling the {\fermi} $\gamma$-ray profiles and spectra
of the Vela pulsar}

We reprocessed the {\fermi} data to obtain the multi-band light
curves in the following steps:
(1) Limited by the timing solution for the Vela
pulsar\footnote{http://fermi.gsfc.nasa.gov/ssc/data/access/lat/ephems/}
from the Fermi Science Support Center (FSSC), we reprocessed the
original data observed from 2008 August 4 to 2009 July 2.
(2) We selected photons of 0.1-300\,GeV in the ``Diffuse" event class,
within a radius of $2^\circ$ of the Vela pulsar position
(RA$=128.55^\circ$, DEC$=-45.75^\circ$) and the zenith angle smaller
than $105^\circ$.
(3) As done by \citet{2009ApJ...696.1084A, 2010ApJ...713..154A,
  2010ApJS..187..460A}, we used ``fselect" to select photons of energy
$E_{\rm GeV}$ within an angle of $<\max[1.6 - 3\log_{10}(E_{\rm
    GeV}),1.3]$ degrees from the pulsar position.
(4) Using the tempo2 \citep{2006MNRAS.369..655H, 2006MNRAS.372.1549E}
with the {\fermi} plug-in, we obtained the rotational phase for each
photon.
(5) Finally we obtained the multi-band $\gamma$-ray light curves with
256 bins, as presented in Figure \ref{GLC} (red solid lines). Two
sharp peaks have a phase separation of $\delta \phi \sim 0.42$. The
ratio of P2/P1 increases with energy. A third broad peak appears in
the bridge emission. The intensity and phase location of P3 vary with
energy.

These observed features challenge all current high energy emission
models. A convincing model with reasonable input parameters for
magnetic inclination angle $\alpha$ and viewing angle $\zeta$ should
produce multi-band light curves of the Vela pulsar and explain the
energy-dependent location of P3 as well as the ratio of P2/P1.

\subsection{Geometric Modeling the Light Curves}

Model parameters for both the annular gap and core gap of the Vela
pulsar should be adjusted for the particle acceleration regions where
the $\gamma$-ray emission are generated. The framework of the annular
gap model as well as the coordinate details have been presented in
\citet{DQHLX+10}, which can be used for simulation of the multi-band
$\gamma$-ray light curves of the Vela pulsar. In this paper, we added
the simulations for the core gap to explain P3 and bridge emission.
We adopted the inclination angle of $\alpha=70^{\circ}$ and the
viewing angle $\zeta=64^{\circ}$ which were obtained from the X-ray
torus fitting \citep{NG08}. The modeling was done as follows.

1. We first separate the polar cap region into the annular and core
gap regions by the critical field line. Then, we use the so-called
``open volume coordinates" ($r_{\rm OVC}, \psi_{\rm s}$) to label the
open field lines for the annular gap and core gap, respectively. Here
$r_{\rm OVC}$ is the normalized magnetic colatitude and $\psi_{\rm s}$
is the magnetic azimuthal. We define $\psi_{\rm s}=0$ for the plane of
the magnetic axis and the spin axis, shown in Figure 1.  For the
annular gap, we define the inner rim $r_{\rm OVC,\, AG} \equiv 0$ for
the critical field lines and the outer rim $r_{\rm OVC,\, AG} \equiv
1$ for the last open field lines; while for the core gap, we define
the outer rim $r_{\rm OVC,\, CG} \equiv 1$ for the critical field
lines and the inner rim $r_{\rm OVC, \, CG} \equiv 0$ for the magnetic
axis. We also divide both the annular gap ($0 \lesssim r_{\rm OVC, \,
  AG} \lesssim 1$) and the core gap ($0.1 \lesssim r_{\rm OVC, \,CG}
\lesssim 1$) into 40 rings for calculation.
\begin{figure}[bt]
\centering
\includegraphics[angle=0,scale=.56]{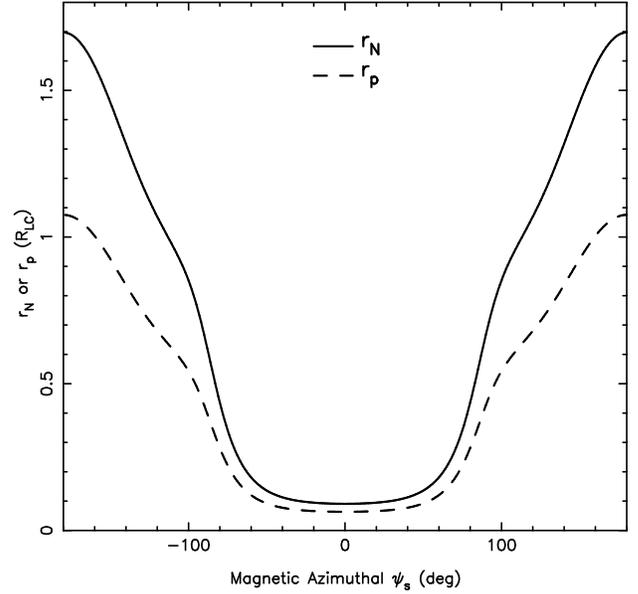}
\caption{The height of the emission peak $r_{\rm p}$ of the Vela
  pulsar in the annular gap model and the height of null charge
  surface $r_{\rm N}$, calculated with an incliantion angle
  $\alpha=70^\circ$, $\kappa=0.7$ and $\lambda=0.9$. Note that $r_{\rm
    p}$ and $r_{\rm N}$ are symmetric around the magnetic axis in the
  magnetic frame. The projected $r_{\rm p}$ is always within the light
  cylinder. We define $\psi_{\rm s}=0^\circ$ for the median between
  the magnetic axis and the equator in the plane of the spin axis and
  magnetic axis.}
\label{RNP}
\end{figure}
\begin{figure*}
\centering
\includegraphics[angle=0,scale=.7]{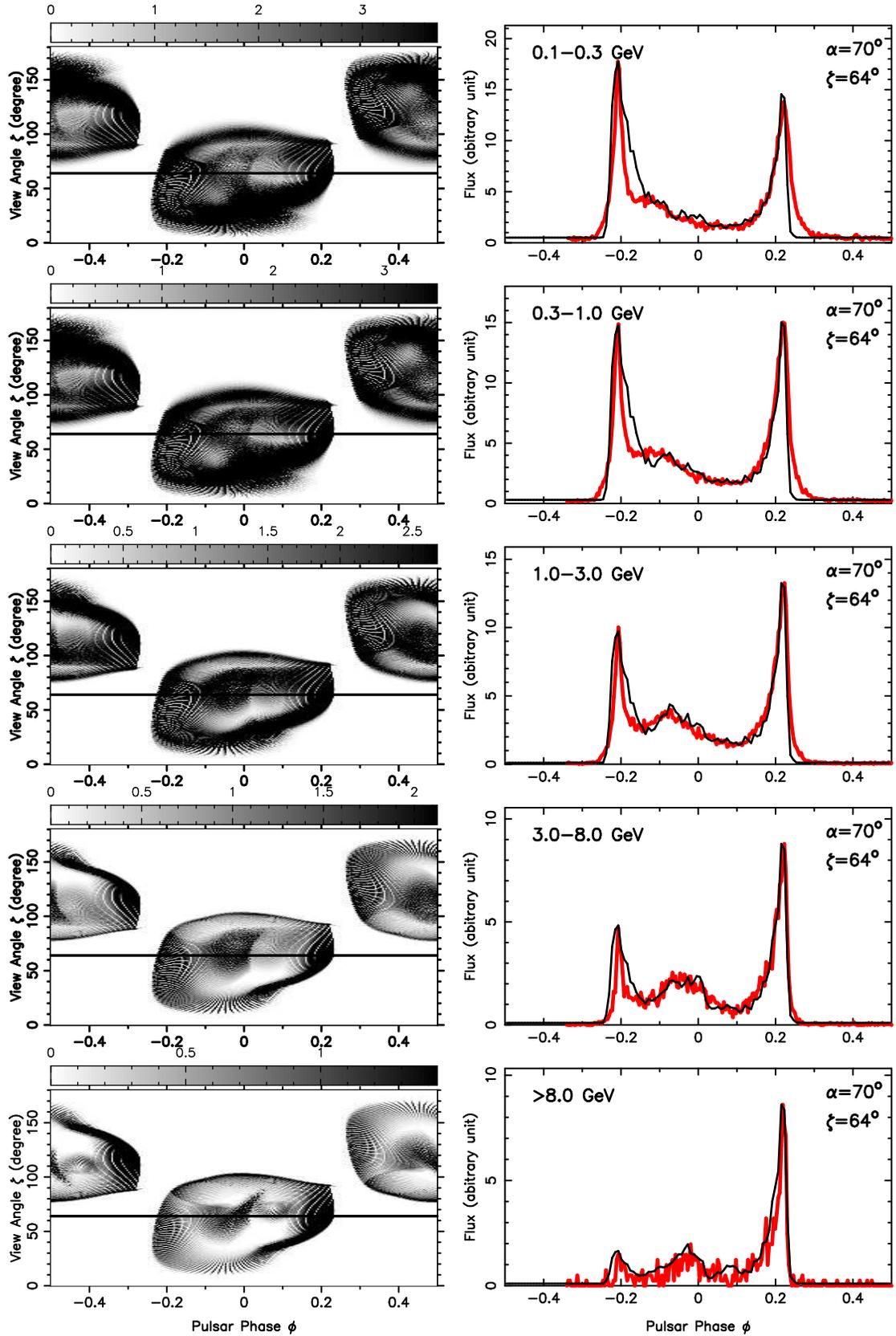}
\caption{The observed multi-band $\gamma$-ray light curves of the Vela
  pulsar (thick red lines in the right panels) and simulated profiles
  from our model (thin black lines). P1 and P2 comes from the annular
  gap region, and P3 and bridge emission from the core gap region. The
  inclination angle $\alpha=70^{\circ}$ and the viewing angle
  $\zeta=64^{\circ}$ (Ng \& Romani 2008) were used for modeling. The
  line of sight cuts across the $\gamma$-ray emission produced from
  only one pole. [{\it See the electronic edition of the Journal
      for a color version of this figure.}]}
\label{GLC}
\end{figure*}

2. Rather following the conventional assumption of the uniform
emissivity along an open field line when modeling the light curves
\citep{2003ApJ...598.1201D,2008ApJ...680.1378H,2010ApJ...709..605F},
for both the annular gap and the core gap, we assume that the
$\gamma$-ray emissivities $I(\theta_{\rm s}, \psi_{\rm s})$ along one
open field line have a Gaussian distribution, i.e.,
\begin{equation} 
I(\theta_{\rm s}, \psi_{\rm s}) =
I_{\rm P}(\theta_{\rm p},\psi_{\rm s}) \exp{\Bigg [
    -\frac{\Big(C(\theta_{\rm s}, \psi_{\rm s})-C_{\rm 0}(\theta_{\rm p},
    \psi_{\rm s}) \Big)^2} {2\sigma_{\rm A}^2}  \Bigg]},
\label{lgs}
\end{equation}
here $\theta_{\rm s}$ is the magnetic colatitude of a spot on a field
line, $\psi_{\rm s}$ is the magnetic azimuthal of this field line,
$C(\theta_{\rm s}, \psi_{\rm s}) = \int_0 ^{\theta_{\rm s}}
\sqrt{r^2+({\rm d}r/{\rm d \theta})^2}\,\rm d \theta $
is the arc length of the emission point on each field line counted
from the pulsar center, $\sigma_{\rm A}$ is a length scale for the
emission region on each open field line in the annular gap or the core
gap in units of $R_{\rm LC}$, and $C_{\rm 0}(\theta_{\rm p}, \psi_{\rm
  s})$ is the arc length for the peak emissivity spot P($\theta_{\rm
  p}, \psi_{\rm s}$) on this open field line. In principle, the peak
position P($\theta_{\rm p}, \psi_{\rm s}$) is dependent on the
acceleration electric field and the emission mechanism. Based on our
1-D calculation of the acceleration field (see Figures \ref{ACC}) and
later the emissivity (see Figure~\ref{PHY} later), the peak emission
comes near the null charge surface.
The height $r_{\rm p, AG}$ for emission peak on open field lines can
be related to the height of the null charge surface
$r_{\rm N}(\psi_{\rm s})$ by
\begin{equation}
r_{\rm p,\, AG}(\psi_{\rm s})=
\lambda \kappa r_{\rm N}(\psi_{\rm  s})+(1-\lambda)\kappa r_{\rm N}(0),
\label{hpa}
\end{equation}
where $\kappa$ is a model parameter for the ratio of heights, and
$\lambda$ is a model parameter describing the deformation of emission
location from a circle \citep[see details in][]{2006AdSpR..37.1988L}.
The emission peak position `P' on each field line can be
uniquely determined, i.e.,
$\theta_{\rm p}=\arcsin [
{\sqrt{r_{\rm p,\, AG}/R_{\rm e, f}(\alpha, \psi_{\rm s})}}
]
$,
where $R_{\rm e, f}(\alpha, \psi_{\rm s})$ is the field line
constant of the open field line with $\psi_{\rm s}$.
Figure \ref{RNP} shows the variations of $r_{\rm p,\, AG}\,(\psi_{\rm
  s})$ and $r_{\rm N}\,(\psi_{\rm s})$ with $\psi_{\rm s}$. The
minimum is at $\psi_{\rm s}=0^\circ$ near the equator and the maximum
at $\psi_{\rm s} = \pm 180^\circ$ near the rotation axis.

The peak emissivity $I_{\rm p}(\theta_{\rm P}, \psi_{\rm s})$ may
follow another Gaussian distribution against $\theta$ for a bunch of
open field lines \citep{2000ApJ...537..964C, 2003ApJ...598.1201D,
  2010ApJ...709..605F}, i.e.,
\begin{equation}
I_{\rm P}(\theta_{\rm p}, \psi_{\rm s}) =
I_{\rm 0} \exp{\Bigg [ -\frac{\Big (\theta_{\rm sp}(\psi_{\rm s})-
      \theta_{\rm cp}(\psi_{\rm s}) \Big)^2} {2\sigma_{\rm \theta}^2} \Bigg ] },
\end{equation}
where $I_{\rm 0}$ is a scaled emissivity, $\sigma_{\rm \theta}$ is a
bunch scale of $\theta$ (in units of rad) for a set of field lines of
the same $\psi_{\rm s}$. $\theta_{\rm sp}$ is used to label a field
line in the pulsar annular regions, $\theta_{\rm cp}=(\theta_{\rm N,
  \psi_{\rm s}}+\theta_{\rm p, \psi_{\rm s}})/2$ (i.e. $r_{\rm
  ovc}\,(\psi_{\rm s}) = 0.5$) is the central field line among those
field lines with $\psi_{\rm s}$.

As seen above, we use two different Gaussian distributions to describe
the emissivity on open field lines for both the annular gap and the
core gap. The model parameters are independently adjusted to maximally
fit the observed $\gamma$-ray light curves. In the core gap, we assume
that the height of emission peak $r_{\rm p,\, CG}=\varepsilon \cdot
r_{\rm p,\, AG} $, where $\varepsilon$ is a model parameter. We
adopted two different $\sigma_{\rm \theta, C}$ for the core gap
because of the different acceleration efficiencies for field lines in
the two ranges of $\psi_{\rm s}$.  We will write $\sigma_{\rm
  \theta,\,A}$ for the annular region and $\sigma_{\rm \theta,\,C}$
for the core region.

3. To derive the ``photon sky-map" in the observer frame, we first
calculate the emission direction of each emission spot ${\bf n_{\rm
    B}}$ in the magnetic frame; then use a transformation matrix
$T_{\rm \alpha}$ to transform ${\bf n_{\rm B}}$ into ${\bf n_{\rm
    spin}}$ in the spin frame; finally use an aberration matrix to
transform ${\bf n_{\rm spin}}$ to ${\bf n_{\rm observer}} = \{ {\bf
  n_{\rm x}, \, n_{\rm y}, \, n_{\rm z} }\}$ in the observer
frame. Here $\phi_{\rm 0} = \arctan ({\bf n_{\rm y}}/{\bf n_{\rm x}})$
and $\zeta = \arccos({\bf n_{\rm z}}/\sqrt{ {\bf n_{\rm x}}^2 + {\bf
    n_{\rm y}}^2 + {\bf n_{\rm z}}^2})$ are the rotation phase
(without retardation effect) of the emission spot with respect to the
pulsar rotation axis and the viewing angle for a distant, nonrotating
observer. The detailed calculations for the aberration effect can be
found in \citet{LD10}.

4. We add the phase shift $\delta \phi_{\rm ret}$ caused by the
retardation effect, so that the emission phase is $\phi=\phi_{\rm
  0}-\delta \phi_{\rm ret}$. Here is no minus sign for $\phi_{\rm 0}$
beacause of the different coordinate systems between our model and the
outer gap model \citep{1995ApJ...438..314R}.
\begin{table}[bt]
\raggedright
\caption{Model parameters for multi-band $\gamma$-ray light curves
 of the Vela pulsar
\label{tbl_1}}
{\footnotesize 
\begin{tabular}{lcccccccc}
\hline 
GeV band & $\kappa$ & $ \lambda$ & $\epsilon$ &
$\sigma_{\rm A}$ & $\sigma_{\rm \theta,\,A}$\tablenotemark{a} &
$\sigma_{\rm \theta,\,C1}$\tablenotemark{b} & $\sigma_{\rm
  \theta,\,C2}$\tablenotemark{c} \\
\hline
0.1--0.3 & 0.68  & 0.9 & 1.17 & 0.5  & 0.0035  & 0.0053  & 0.009 \\
0.3--1.0 & 0.70  & 0.9 & 1.20 & 0.5  & 0.0035  & 0.0046  & 0.009  \\
1.0--3.0 & 0.72  & 0.9 & 1.15 & 0.5  & 0.0014  & 0.0064   & 0.01   \\
3.0--8.0 & 0.72  & 0.9 & 0.88 & 0.5  & 0.0007  & 0.0085  & 0.006  \\
$>$8.0  & 0.73  & 0.9 & 0.82 & 0.1  & 0.00085  & 0.007   & 0.003   \\
\hline
\end{tabular}
{}$^{\rm a}$The bunch scale for field lines in the annular gap.\\
{}$^{\rm b}$The bunch scale for field lines of
$90^\circ<\psi_{\rm s}<180^\circ$ in the core gap.\\
{}$^{\rm c}$The bunch scale for field lines of
  $-180^\circ<\psi_{\rm s}<90^\circ$ in the core gap. }

\end{table}

5. The ``photon sky-map'', defined by the binned emission intensities
on the ($\phi$, $\zeta$) plane, can be plotted for 256 bins (see
Figure \ref{GLC}). The corresponding light curves cut by a line of
sight with a viewing angle $\zeta=64^{\circ}$ are therefore finally
obtained. For the viewing angle $\zeta=64^{\circ}$, any magnetic
inclination angles of $\alpha$ between $60^{\circ}$ and $75^{\circ}$
in the annular gap model can produce light curves with two sharp peaks
and a large peak separation (e.g., 0.4 -- 0.5), similar to the
observed ones. The emission from the single pole is favored for the
Vela pulsar in our model.

The modeled light curves are presented in Figure \ref{GLC} (black
solid lines), with the model parameters listed in Table
\ref{tbl_1}. Emission of P1 and P2 comes from the annular gap region
in the vicinity of the null charge surface, and P3 and bridge emission
comes from the core gap region. The higher energy P3 emission
($>3$\,GeV) comes from lower height, whereas the lower energy
$\gamma$-ray emission comes from a higher region. In the annular gap
region, higher energy emission is mostly generated in higher region.
Nevertheless, the $\gamma$-ray emission heights are above the lower
bound of the height determined by $\gamma-B$ absorption \citep{LD10}.

The peak emission comes from different field lines and emission
heights in the annular gap. The deformation of radiation beam is
related to high value of geometric factor $\lambda$ as discussed in
\citet{DQHLX+10}. Owing to the aberration and retardation effects, the
enhanced gamma-ray emission in the outer rim of photon sky-map make
the peak very sharp, especially for P2. For the Vela pulsar, the high
inclination angle of about $\alpha=70^{\circ}$ is important to get the
observed two sharp peaks with a large separation.
\begin{figure}[b]
\centering
\includegraphics[angle=0,scale=.54]{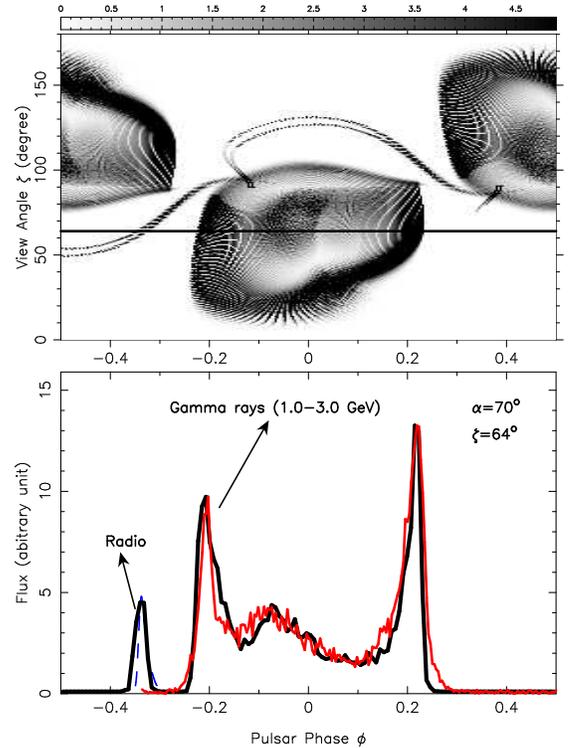}
\caption{The jointly modeled radio and $\gamma$-ray light curves
  (black solid lines) for the Vela pulsar. The radio emission is
  prodeuced from a higher and narrower region in the annular gap
  region of the same magnetic pole as the $\gamma$-ray. The observed
  radio profile (blue dashed line) for the Vela pulsar is taken from
  the website http://fermi.gsfc.nasa.gov/ssc/data/access/lat/ephems/,
  and the $\gamma$-ray profile is observed by the {\fermi}. [{\it See
      the electronic edition of the Journal for a color version of
      this figure.}]  } \label{LAG}
\end{figure}

\subsection{Radio Lag}

With well-coordinated efforts for pulsar timing program,
\citet{2010ApJS..187..460A} determined the phase lag between radio
emission and $\gamma$-ray light curves. The radio pulse comes earlier
by a phase of $\sim 0.13$ (see Figure \ref{LAG}).

Radio emission might be generated in the two locations of a pulsar
magnetosphere. One is the traditional low-height polar cap region for
long-period ($P \sim 1 $s) pulsars \citep{RS75}. The other is the
outer magnetospheric region with high altitudes near the light
cylinder \citep{2005ApSS.297..101M}. For the polar cap region, the low
radio emission height leads to a small beam, which probably does not
point to an observer for the Vela pulsar. \citet{2010ApJ...716L..85R}
propose that the radio emission from young pulsars is radiated in a
high region close to the null-charge surface, i.e. the similar region
for $\gamma$-ray emission. This is somehow similar to our annular gap
model, in which the radio emission originates from a higher and
narrower region than that of the $\gamma$-ray emission.

The modeled radio and $\gamma$-ray light curves in the two-pole
annular gap model are shown in Figure \ref{LAG}. The region for the
radio emission is mainly located at a height of $\sim R_{\rm LC}$ on
certain filed lines with $\psi_s=-138^{\circ}$. Our scenario of radio
emission for the Vela pulsar is consistent with the narrow stream of
hollow-cone-like radio emission \citep{2010MNRAS.401.1781D}. According
to our model, not all $\gamma$-ray pulsars can be detected in the
radio band, and not all radio pulsars can have a $\gamma$-ray beam
towards us.

\subsection{$\gamma$-ray Spectra for the Vela Pulsar}

\citet{2010ApJ...713..154A} got high quality phase-resolved spectra
(P1, P2, low-energy P3 and high-energy P3) and the phase-averaged
spectrum of the Vela pulsar. The observed $\gamma$-ray emission is
believed to originate from the curvature radiation of primary
particles \citep{2008ApJ...676..562T, 2008ApJ...680.1378H,
  Meng08}. Here we use the synchro-curvature radiation from primary
particles \citep{1995PhLA..208...47Z, 1996ApJ...463..271C, Meng08} and
also the synchrotron radiation from secondary particles to calculate
the $\gamma$-ray phase-averaged and phase-resolved spectra of the Vela
pulsar.

We divide the annular gap region into 40 rings and 360 equal intervals
in the magnetic azimuth, i.e. in total 40$\times$360 small magnetic
tubes. A small magnetic tube has a small area $A_0$ on the neutron
star surface. From equation (\ref{rho}), the number density of primary
particles at a height $r$ is $n(r) = \frac{\Omega B(r)}{2 \pi c\; e}
\cos\zeta_{\rm out}$, where $c$ is the speed of light, and $e$ is the
electric charge. The cross-section area of the magnetic tube at $r$ is
$A(r)=B_0A_0/B(r)$. Therefore, the flowing particle number at $r$ in
the magnetic tube is
\begin{equation}
\Delta N(r) =
A(r)\Delta s\frac{\Omega B(r)}{2 \pi c e}\cos\zeta_{\rm  out} ,
\label{dnf}
\end{equation}
here $\Delta s$ is the arc length along the field.

The accelerated particles are assumed to flow along a field line in a
quasi-steady state. Using the calculated acceleration electric field
shown in Figure~\ref{ACC}, we can obtain the Lorentz factor $\gamma$
of the primary particle from the curvature radiation reaction
\begin{equation}
\gamma = (\frac{3\rho^2 E_{\parallel}}{2e})^{\frac{1}{4}} = 2.36\times
  10^7{\rho_7}^{0.5} E_{\parallel,\, 6}^{0.25},
\label{gam}
\end{equation}
where $\rho_7$ is the curvature radius in units of $10^7$\,cm and
$E_{\parallel,\, 6}$ is the acceleration electric field in units of
$10^6 \rm V\, cm^{-1}$.
The pitch angle $\beta$ of the primary particles flowing along
a magnetic field line is \citep{Meng08}
\begin{equation}
\sin\beta \approx \beta \approx \eta \frac
{\gamma m_{\rm e}c^2}{eB(r)\rho},
\label{beta}
\end{equation}
where $\eta \leq 1$,  $m_{\rm e}$ is the electron mass, and $\rho$ is the
curvature radius.
The characteristic energy $E_{\rm c}^{\rm syn-cur}$
of synchro-curvature radiation \citep{1995PhLA..208...47Z, Meng08} is
given by
\begin{eqnarray}
& E&_{\rm c}  ^{\rm syn-cur} = \frac{3}{2}\hbar c \gamma^3\frac{1}{\rho} \\ \nonumber
&\times& \sqrt{ (\frac{r_{\rm B}}{\rho}+1-3\frac{\rho}{r_{\rm B}} )\cos^4\beta +
    3\frac{\rho}{r_{\rm B}} \cos^2\beta +\frac{\rho^2}{r_{\rm B}^2} \sin^4\beta } ,
\label{ecc}
\end{eqnarray}
where $r_{\rm B} = \frac{\gamma m_{\rm e}c^2 \sin\beta}{eB(r)}$ is the
cyclotron radius of an electron, and $\hbar$ is the reduced Planck
constant.

The energy spectrum ${\rm d} N/{\rm d} \gamma$ of the accelerated
primary particles is unknown. \citet{2008ApJ...680.1378H} have assumed
it to follow a broken power-law distribution for pairs with indexes of
$-2.0$ and $-2.8$ [see their equation (47)]. Here we assume the
primary particles in the magnetic tube to follow one power law ${\rm
  d}N/{\rm d}\gamma = N_{\rm 0}\gamma^{\Gamma}$ with an index of
$\Gamma=-2.4$. Here, $N_{\rm 0}$ can be derived by integration the
equation above using the equations (\ref{dnf}) and (\ref{gam}). The
$\gamma$-ray spectrum emitted by the primary particle can be
calculated by \citep{Meng08}
\begin{eqnarray}
& F & (E_\gamma) = \frac{E_{\rm \gamma}^2}{\Delta \Omega d^2}
    \frac{{\rm d}^2 N_\gamma}{{\rm d}E_\gamma {\rm d}t } =
    \frac{\sqrt{3}e^2}{2 h \Delta \Omega d^2}
    \int_{\gamma_{\rm min}}^{\gamma_{\rm max}} \frac {{\rm d} N} {{\rm
        d} \gamma} E_{\gamma} \\ \nonumber
 & \times & \frac{\gamma}{r_{\rm C}} \Bigg [ \Bigg ( 1 +
  \frac{1}{r_{\rm C}^2 Q_2^2} \Bigg ) x G(x) - \Bigg ( 1 -
  \frac{1}{r_{\rm C}^2 Q_2^2}\Bigg ) x K_{\rm 2/3}(x) \Bigg ] {\rm d} \gamma ,
\end{eqnarray}
where $\Delta\Omega$ is the solid angle of the $\gamma$-ray beam, $h$
is the Planck constant, $x=E_\gamma/E_{\rm c}^{\rm syn-cur}$,
$G(x)=\int_x^{+\infty} K_{\rm 5/3}(z){\rm d}z$, $K_{\rm 5/3}(z)$ and $
K_{\rm 2/3}(x)$ are the modified bessel function with the order of 5/3
and 2/3, and $r_{\rm C}$ and $Q_2^2$ are given by
\begin{displaymath}
r_{\rm C} =
\frac{c^2} {[(r_{\rm B}+\rho)\Omega_0^2 + r_{\rm B}\omega_{\rm B}^2]},
\end{displaymath}

\begin{displaymath}
\Omega_0 = \frac{c \cos\beta}{\rho}, \;\;\;\;
\omega_{\rm B}=\frac{eB(r)}{\gamma m_{\rm e}c},
\end{displaymath}

\begin{displaymath}
{\small Q_2^2 =
\frac{1}{r_{\rm B}}\Bigg( \frac{r_{\rm B}^2+r_{\rm B}\rho-3\rho^2}
{\rho^3}\cos^4\beta + \frac{3}{\rho}\cos^2\beta
  + \frac{1}{r_{\rm B}}\sin^4\beta \Bigg), }
\end{displaymath}
respectively.

\begin{table*}[tb]
\centering
\caption {Best fit parameters for modeling the phase-averaged spectrum of
the Vela pulsar.
\label{tbl_2}}
\begin{tabular}{ccccccccc}
\hline
    & $\psi_{\rm s}\,(^\circ)$ & $R_{\rm e}$ &
$r$ & $\gamma_{\rm min}^{\rm pri}$ & $\gamma_{\rm
  max}^{\rm pri}$ & $\Delta\Omega $ & $\gamma_{\rm min}^{\rm 2nd}$ &
$\gamma_{\rm max}^{\rm 2nd}$  \\
\hline
P1 & -110 & 1.095 & 0.62 & 0.50$\times10^7$ & 1.79$\times10^7$ & 0.11
& 8.45$\times10^5$   &   1.31$\times10^6$           \\
P2 & 131 & 1.278 & 0.75 & 0.30$\times10^7$ & 2.65$\times10^7$ & 1.05
& 6.25$\times10^5$   &   1.75$\times10^6$         \\
P3 & -104 & 1.122 & 0.28 & 0.35$\times10^7$ & 2.40$\times10^7$ & 1.09
& 3.05$\times10^5$   &   1.22$\times10^6$         \\
\hline
\end{tabular}   %
\tablecomments{\centering $R_{\rm e}$ is the field line constant in
  units of $R_{\rm LC}$, and $r$ is the emission height of a profile
  component.}
\end{table*}

\begin{figure*}[tb]
\centering
\includegraphics[angle=0,scale=.53]{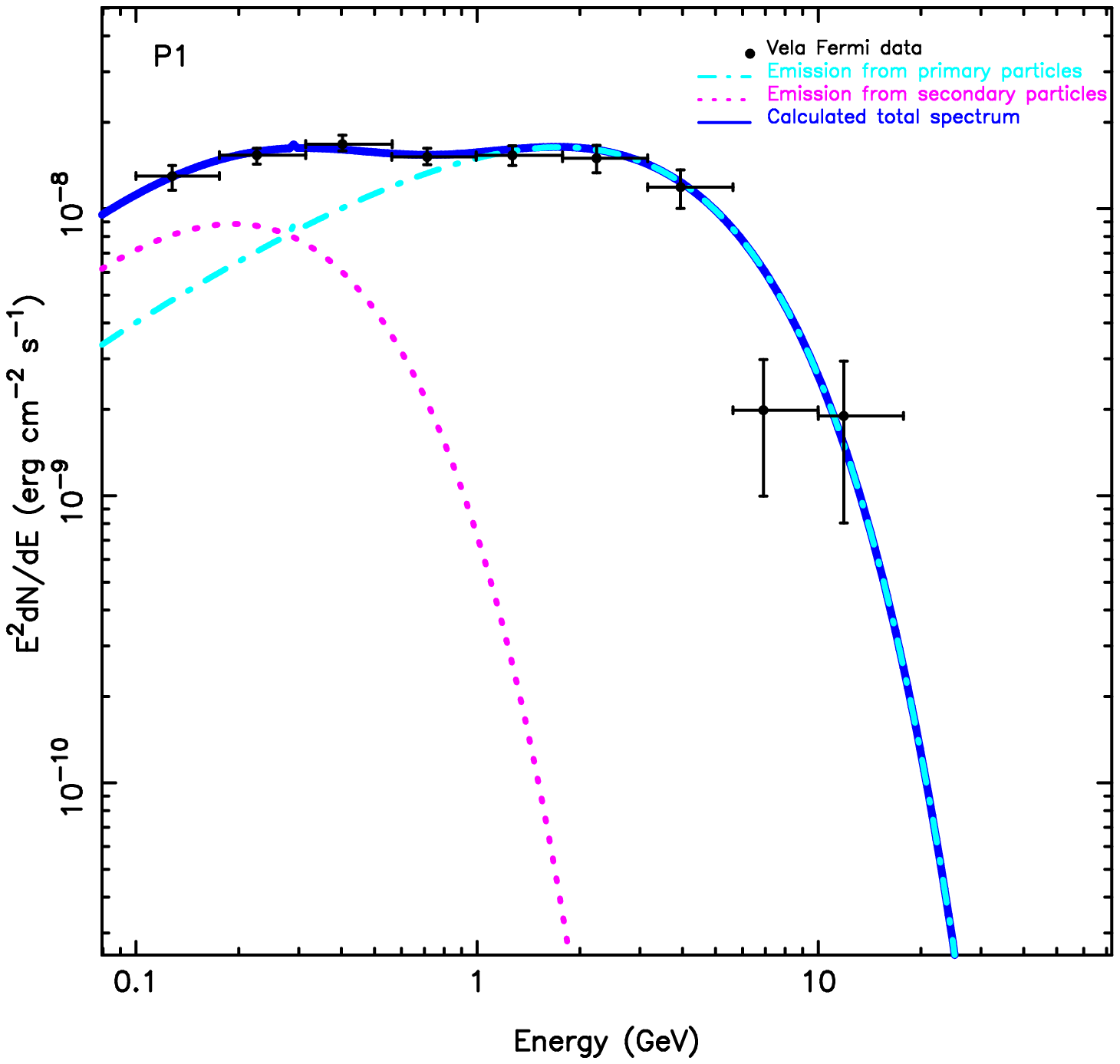}
\includegraphics[angle=0,scale=.53]{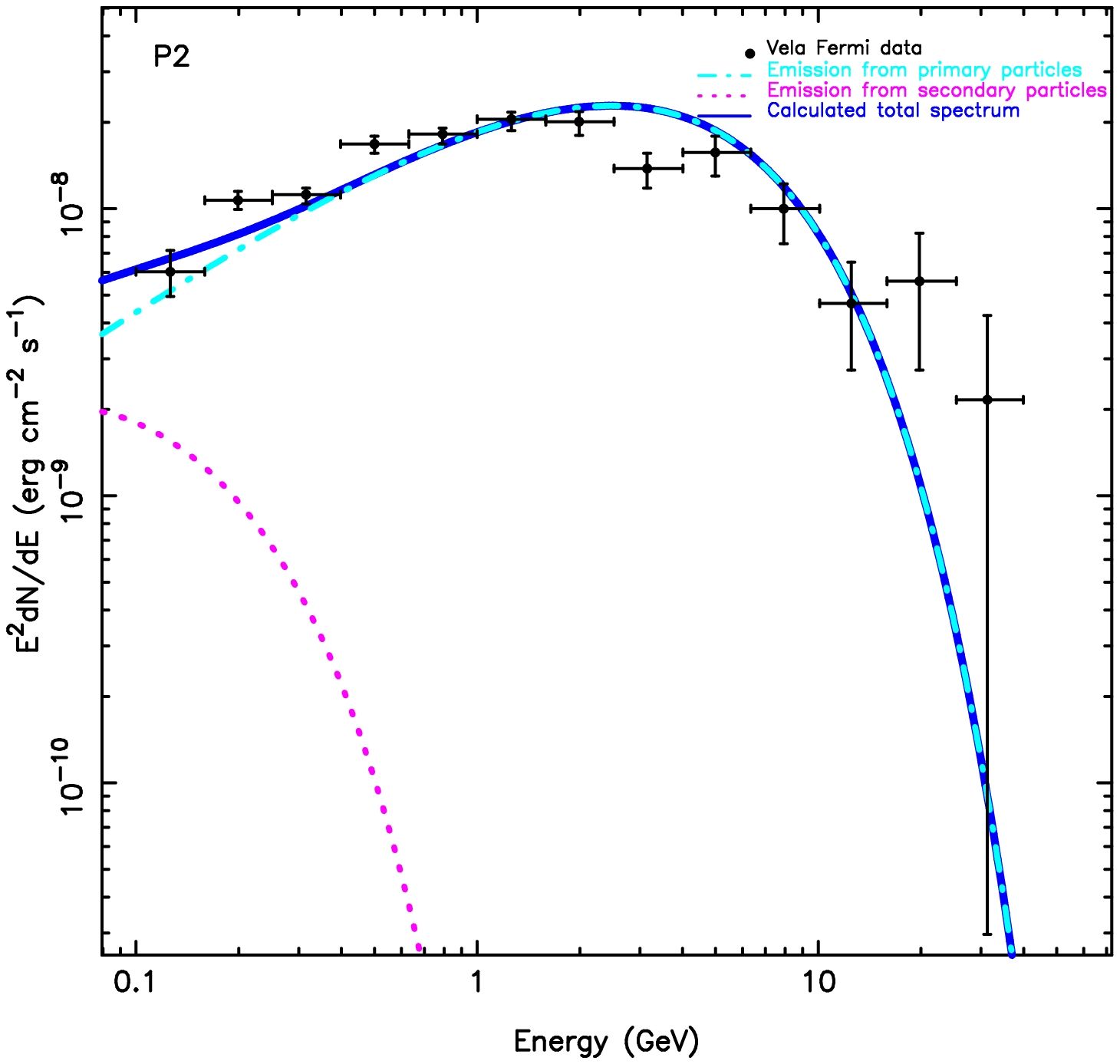}
\includegraphics[angle=0,scale=.53]{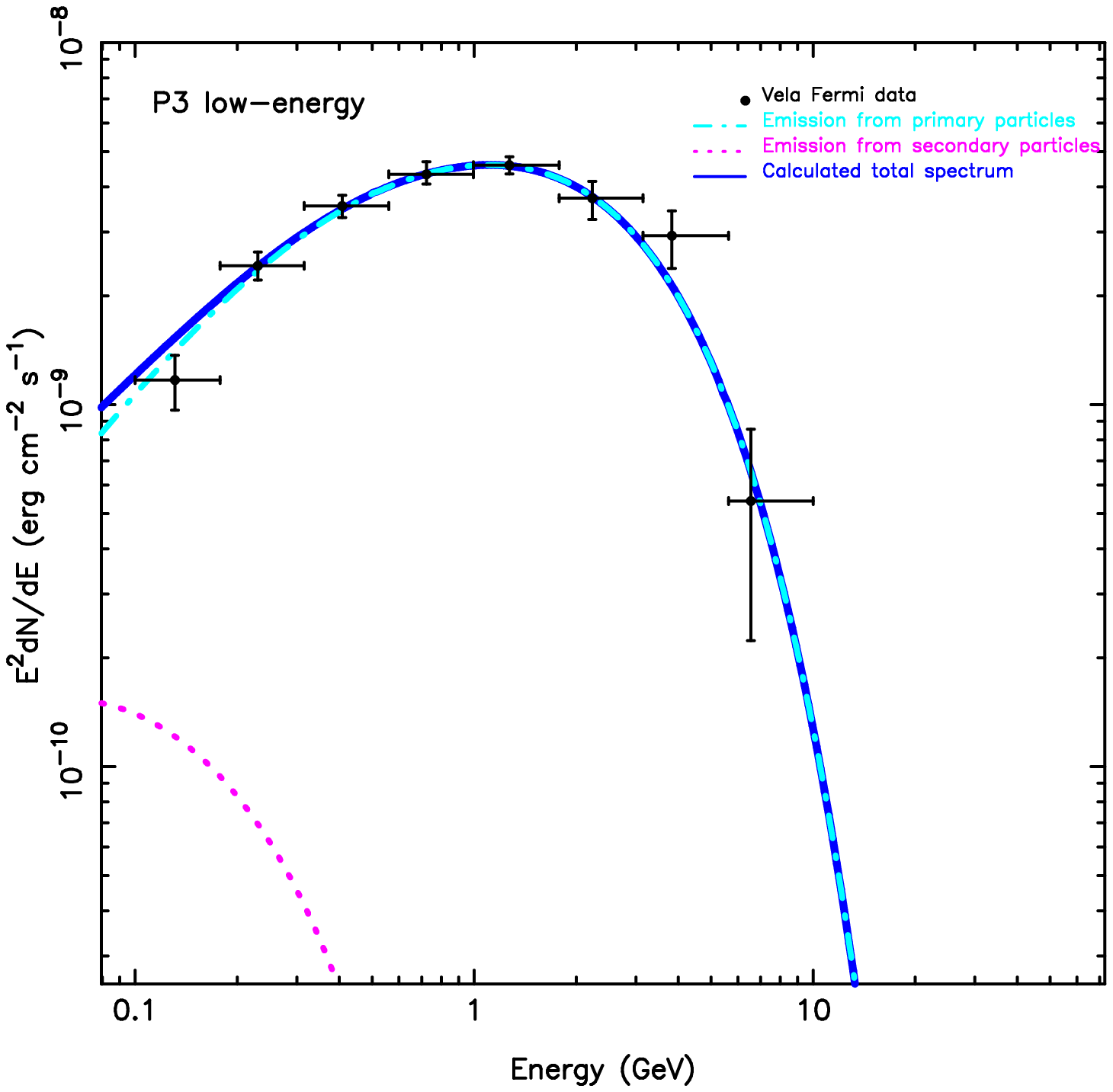}
\includegraphics[angle=0,scale=.53]{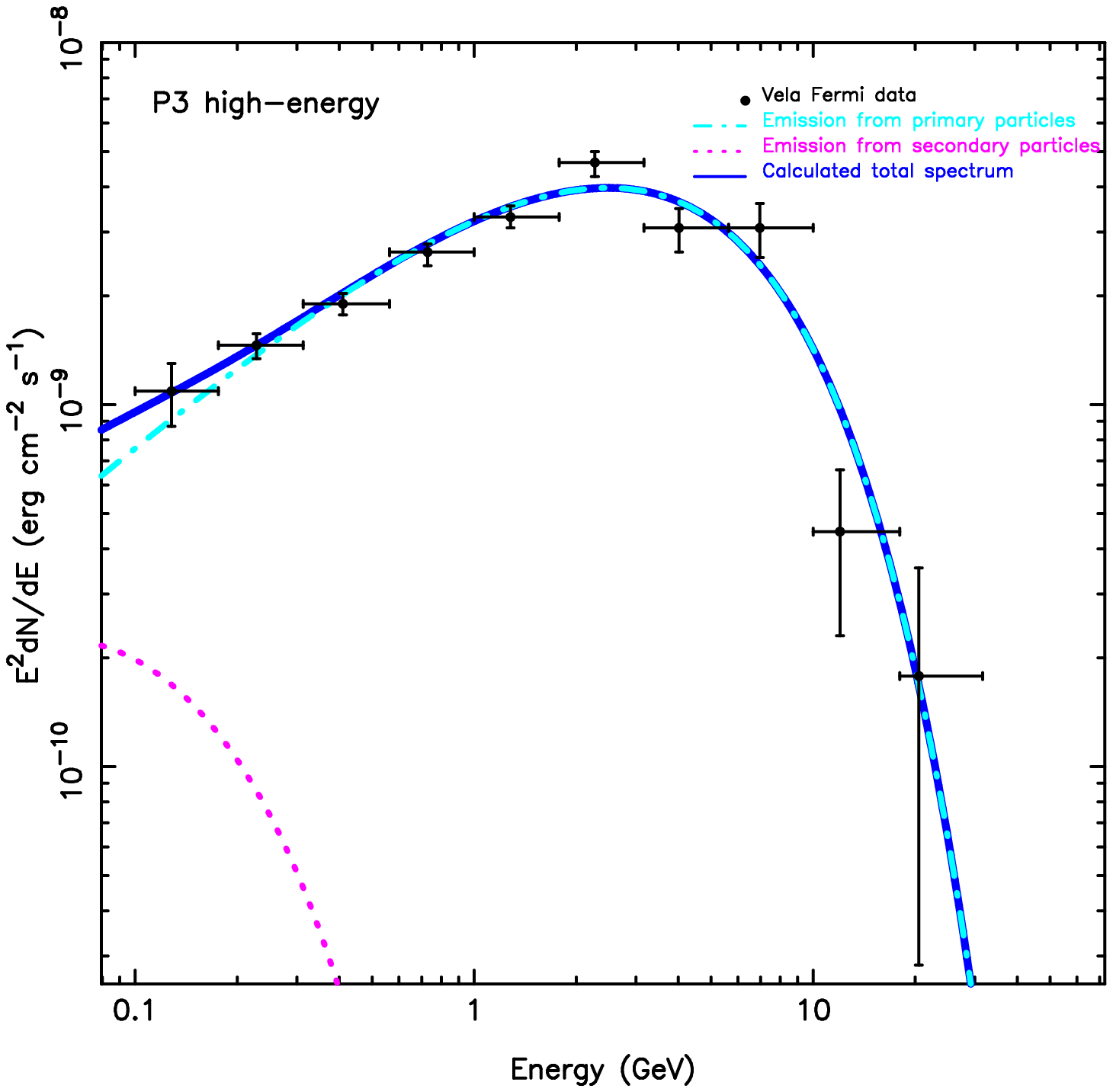}
\caption{Fitting the $\gamma$-ray phase-resolved spectra for the Vela
  pulsar. The observed data were taken from
  \citet{2010ApJ...713..154A} and plotted as black points with
  error-bars. The contributions of $\gamma$-ray emission from the both
  primary particles and secondary particles are plotted separately for
  profile components, P1, P2, P3 low-energy and P3 high-energy. The
  emission from the secondary particles only partly contribute the
  lower energy band.  [{\it See the electronic edition of the Journal
      for a color version of this figure.}]  }
\label{PRS}
\end{figure*}

\begin{figure}
\centering
\includegraphics[angle=0,scale=0.53]{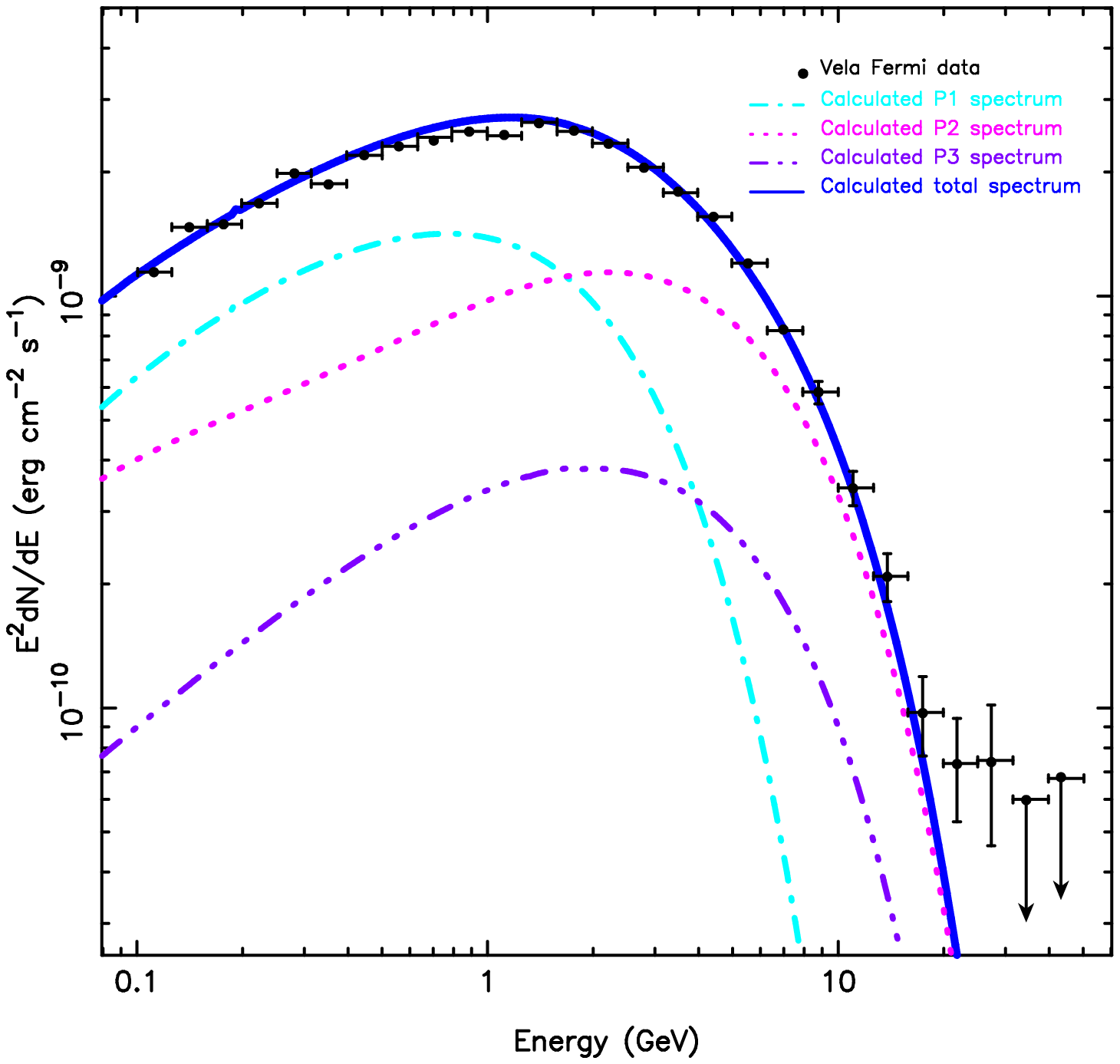}
\caption{Fitting the $\gamma$-ray phase-averaged spectrum for the Vela
  pulsar. The observed data were taken from
  \citet{2010ApJ...713..154A} and plotted as black points with
  error-bars. The spectra for three profile components and the total
  phase-averaged spectrum of the Vela pulsar are modeled from the
  synchro-curvature radiation from primary particles and synchrotron
  radiation from secondary particles. [{\it See
      the electronic edition of the Journal for a color version of
      this figure.}]}
\label{PAS}
\end{figure}

The secondary particles can be generated with a large multiplicity
($10^3 - 10^4$) via the $\gamma-B$ process in the lower regions of the
annular gap and the core gap near the neutron star surface. Here,
we assume that the energy spectrum of secondary particles follow a
power-law, with an index of $\Gamma_{\rm sec}=-2.8$ and a multiplicity
of $M_{\rm sec} \sim 1000$. The pitch angle of pairs increase due
to the cyclotron resonant absorption of the low-energy photons
\citep{2008ApJ...680.1378H}. The mean pitch angle of secondary
particles is about 0.06, adopted from equation (\ref{beta}) with a
slightly large factor $\eta \gtrsim 1$ owing to the effect of
cyclotron resonant absorption. The synchrotron radiation from
secondary particles have some contributions to the low-energy
$\gamma$-ray emission, e.g. $\lesssim 0.3$~GeV.

We further checked the optical depth $\tau_{\rm \gamma-B}$ of the
$\gamma-$B absorption \citep{LD10}
\begin{equation}
\tau_{\rm \gamma-B}\,(r) =
\frac{1.55\times10^7 r}{E_{\rm \gamma}}K_{1/3}^2
(\frac{2.76\times10^6 r^{5/2}P^{1/2}}{B_{\rm 0,\,12}R^3 E_{\rm \gamma}} ),
\label{tau}
\end{equation}
here $E_{\rm \gamma}$ is in units of MeV, $B_{\rm 0,\,12}$ is in units
of $10^{12}$\,G. We found that the {\fermi} $\gamma$-photons of the
Vela pulsar with an energy of $<50$\,GeV always have a $\tau_{\rm
  \gamma-B}\ll 1$ if the emission height is greater than a few hundred
kilometers.

To reduce the computation time, we calculate the synchro-curvature
radiation at the ``averaged emission-height" for three components, P1,
P2 and P3, of the $\gamma$-ray light curve of the Vela pulsar. For P1,
the emission height is about $0.62 R_{\rm LC}$ on the field line of a
magnetic azimuth $\psi=-110^\circ$; for P2, the emission height is
$0.75 R_{\rm LC}$ on the field line of $\psi=131^\circ$; and for P3,
the emission height is about $0.28 R_{\rm LC}$ on the field line of
$\psi=-104^\circ$. We compute $E_\parallel$ for the three peaks, and
adjust the minimum and maximum Lorentz factor for primary particles,
$\gamma_{\rm min}^{\rm pri}$ and $\gamma_{\rm max}^{\rm pri}$, and the
minimum and maximum Lorentz factor for secondary particles,
$\gamma_{\rm min}^{\rm 2nd}$ and $\gamma_{\rm max}^{\rm 2nd}$, and the
$\gamma$-ray beam angle $\Delta\Omega$ to fit the $\gamma$-ray spectra
for the Vela pulsar.

We fitted the phase-averaged spectrum and phase-resolved (P1, P2,
low-energy P3 and high-energy P3) spectra of the Vela pulsar as shown
in Figure~\ref{PRS} and Figure~\ref{PAS}. The best fit parameters for
phase-resolved and phase-averaged spectra are similar as expected, and
are listed in Table \ref{tbl_2}. The maximum Lorentz factor of primary
particles $\gamma_{\rm max}^{\rm pri}$ is consistent with that
obtained from the curvature radiation balance of the outer
magnetosphere models given by \citet{2010ApJ...713..154A}. The modeled
spectra are not sensitive to $\gamma_{\rm min}^{\rm pri}$ or
$\gamma_{\rm min}^{\rm 2nd}$, but quite sensitive to $\gamma_{\rm
  max}^{\rm pri}$ which is chosen around the value of the steady
Lorentz factor given by equation (\ref{gam}).  The solid angle
$\Delta\Omega$ was always assumed to be 1 by many authors for
simplicity. We adjusted it as a free parameter around 1 for different
phases.

The synchro-curvature radiation from primary particles is the main
origin of the observed $\gamma$-ray emission, while the synchrotron
radiation from secondary particles can contribute to the lower energy
band to improve the fitting. The peak ratios of P1 and P2 shown in
Figure \ref{PAS} are roughly consistent with observations except for
the band of 0.3--1.0\,GeV (cf. Figure \ref{GLC}). The high-energy P3
is generated in the relatively low height of the core gap, where the
particles have a higher acceleration efficiency than those for the
low-energy P3, which lead to their cutoff energy different. The
phase-resolved spectra for both high-energy P3 and low-energy P3 can
be explained in the synchro-curvature radiation from primary particles
from the core gap, with little contribution from the synchrotron
radiation of secondary particles because they in general have small
pitch angles with respect to field lines and large curvature
radius. However, the synchrotron radiation from secondary particles
does contribute to the $0.1 - 0.3\,$GeV band for P1 and P2.

\begin{figure}
\centering
\includegraphics[angle=0,scale=.50]{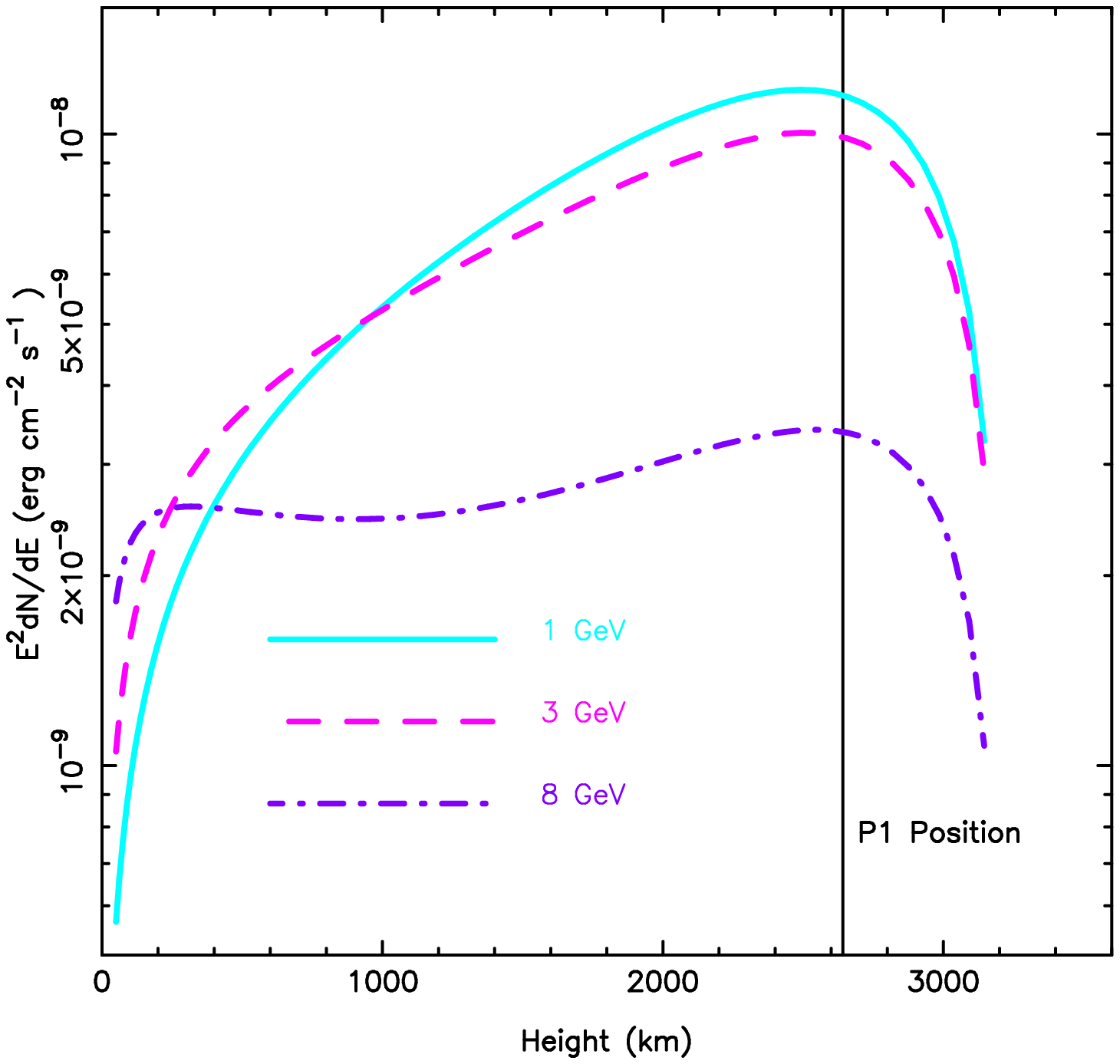}
\includegraphics[angle=0,scale=.50]{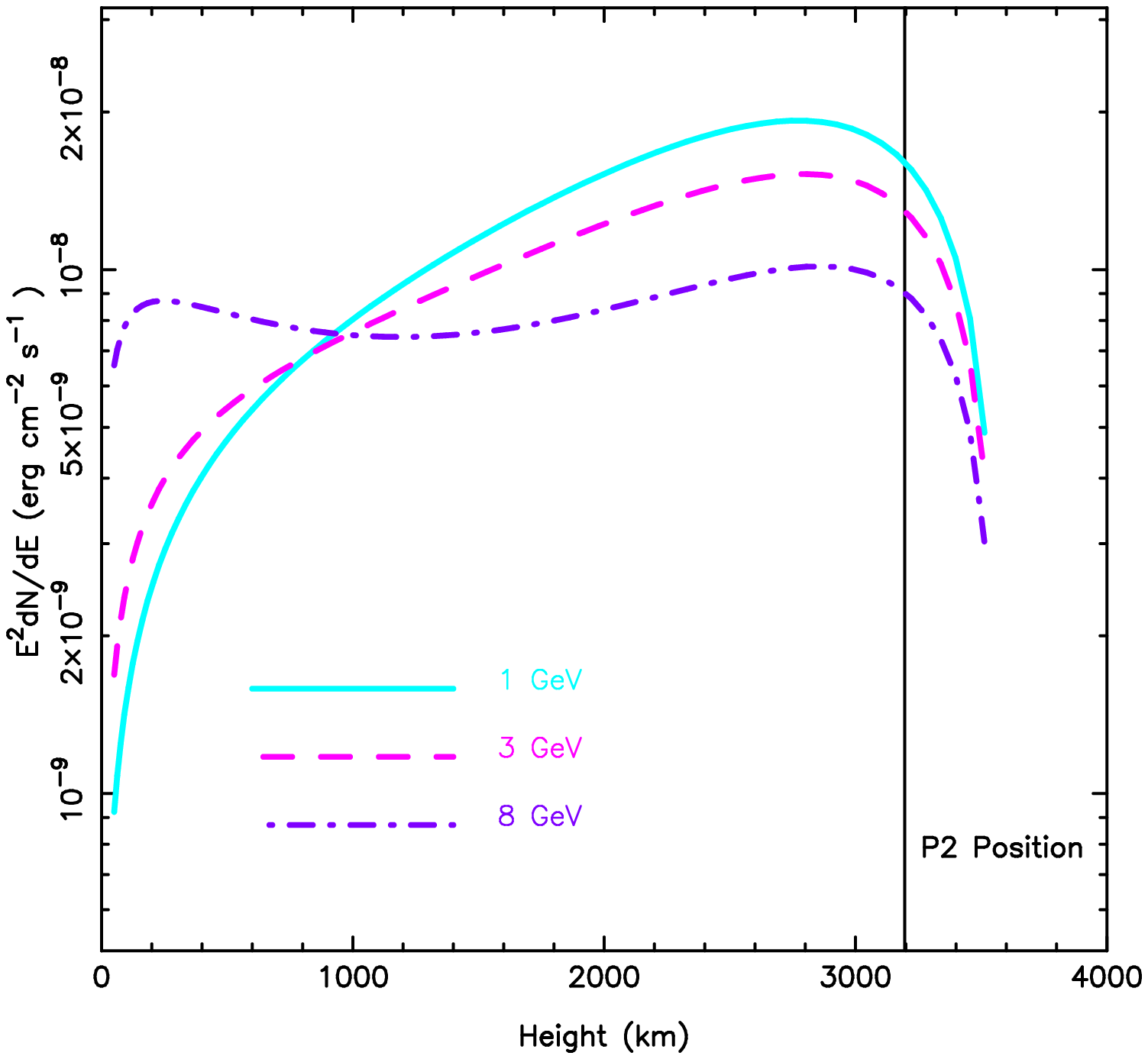}
\caption{The emission at 1\, 3 and 8\,GeV for the P1 ({\it top
    pannel}) and P2 ({\it bottom pannel}) is not uniform along open
  field lines. It varies with the height. [{\it See the electronic
      edition of the Journal for a color version of this figure.}] }
\label{PHY}
\end{figure}

In Figure \ref{PHY}, we plot the emission fluxes of P1 and P2
component at 1, 3 and 8\,GeV of the Vela pulsar against the emission
height. It is not uniform along an open field line. The bump at a low
height for high energy $\gamma$-ray (e.g. $\gtrsim$8 GeV) due mainly
to the small curvature radius and large acceleration electric field
there.  In Section 3.1, we roughly took a Gaussian distribution along
the arc (equation \ref{lgs}) to decribe the emissivity near the peak
emission region, which is natural in our annular gap model and
independent of the model paranmeters.

\section{Discussions and Conclusions}

The detailed features of $\gamma$-ray pulsed emission of the Vela
pulsar observed by {\fermi} provide challenge to current emission
models for pulsars.

The charged particles can not co-rotate with a neutron star near the
light cylinder, and must flow out from the magnetosphere. To keep the
whole system charge-free, the neutron star surface must have the
charged particles flowing into the magnetosphere. We found that the
acceleration electric field $E_\parallel$ in a pulsar magnetosphere is
strongly correlated with the GJ density $\rho_{\rm gj}$ near the light
cylinder radius $R_{\rm LC}$, while $\rho_{\rm gj}$ at $(R_{\rm LC})$
is proportional to the local magnetic field $B_{\rm LC}$. It has been
found that the {\fermi} $\gamma$-ray pulsars can be young pulsars and
millisecond pulsars which have a high $B_{\rm LC}$. This means that
the acceleration electric field $E_\parallel$ in a pulsar magnetosphere
is related to the observed {\fermi} $\gamma$-ray emission from pulsars.

To well understand the multi-band pulsed $\gamma$-ray emission from
pulsars, we considered the magnetic field configuration and 3-D global
accleration electric field with proper boundary conditions for the
annular gap and the core gap. We developed the 3D annular gap model
combined with a core gap to fit the $\gamma$-ray light curves and
spectra. Our results can reproduce the main observed features for the
Vela pulsar. The emission peaks, P1 and P2, originate from the annular
gap region, and the P3 and bridge emission comes from the core gap
region.  The location and intensity of P3 are related to the emission
height in the core region. The higher energy emission ($>3$\,GeV)
comes from lower regions below the null charge surface, while the
emission of lower energy of less than 3\,GeV comes from the region
near or above the null charge surface. Radio emission originates from
a region, higher and narrower than those for the $\gamma$-ray
emission, which explains the phase lag of $\sim 0.13$ prior to P1,
consistent with the model proposed by \citet{2010MNRAS.401.1781D}.

Synchro-curvature radiation is a effective mechanism for charged
particles to radiate in the generally curved magnetic field lines in
pulsar magnetosphere \citep{1995PhLA..208...47Z,
  1996ApJ...463..271C}. The GeV band emission from pulsars is
originated mainly from curvature radiation from primary particles,
while synchrotron radiation from secondary particles have some
contributions to the low-energy $\gamma$-ray band (e.g., $0.1 -
0.3$~GeV). Moreover, contributions of curvature radiation from
secondary particles and inverse Compton scattering from both primary
particles and secondary particles could be ignored in the $\gamma$-ray
band. The synchro-curvature radiation from the primary particles and
synchrotron radiation from secondary particles are calculated to model
the phase-resolved spectra for P1, P2 and P3 of low-energy band and
high-energy band and the total phase-averaged $\gamma$-ray spectrum.

In short, the $\gamma$-ray emission from the Vela pulsar can be well
modeled with the annular gap and core gap.

\acknowledgments
The authors are very grateful to the referee and Dr. Wang Wei for
helpful comments. YJD thanks the COSPAR community for the final
support to participate the 11th COSPAR Capacity-Building Workshop on
"Data Analysis of the Fermi Gamma-ray Space Telescope" held in
Bangalore, India during 2010 February 8 to February 19. He also thanks
Professor Biswajit Paul and the Raman Research Institute for kind
helps, and Professor Thompson, D.~J. for fruitful discussions and the
tempo2 {\fermi} plug-in.  We also thank both the pulsar groups of NAOC
and of Peking University for useful conversations. The authors are
supported by NSFC (10821061, 10573002, 10778611, 10773016 and
10833003) and the Key Grant Project of Chinese Ministry of Education
(305001).

\end {document}